\documentclass[aps,amssymb,prl, amsmath, twocolumn, floatfix]{revtex4}
\usepackage{graphics}
\usepackage{psfrag}
\usepackage{epsfig}
\usepackage{subfigure}
\begin{document}
\bibliographystyle{prsty}

\title{Force networks and the dynamic approach to jamming in sheared granular media}
\author{Gregg Lois}
\author{Jean M. Carlson}
\affiliation{
Department of Physics, University of California, Santa Barbara, California 93106}

\begin{abstract}
Diverging correlation lengths on either side of the jamming transition are used to formulate a rheological model of granular shear flow, based on the propagation of stress through force chain networks.
The model predicts three distinct flow regimes, characterized by the shear rate dependence of the stress tensor, that have been observed in both simulations and experiments.  The boundaries separating the flow regimes are quantitatively determined and testable.  In the limit of jammed granular solids, the model predicts the observed anomalous scaling of the shear modulus and a new relation for the shear strain at yield.
\end{abstract}
\maketitle

Jamming takes place when an amorphous collection of particles spontaneously develops rigidity and supports weight like a solid instead of flowing like a liquid~\cite{liujamming}.
The transition occurs without static spatial ordering, but is accompanied by long range dynamical correlations arising from the collective motion of groups of particles~\cite{gajkinetic, silbertwyart, others}.
In the case of sheared granular materials, the transition from jammed to flowing phases occurs as the applied stress ${\bf \Sigma}$ is increased or the packing fraction $\phi$ is decreased, and there is a well defined packing fraction $\phi_\mathrm{c}$ at which the system jams in the limit of zero stress~\cite{ohern, zhangmakse}.
% applied stress and the packing fraction, as illustrated in Figure~\ref{jammingdiagram}.  
Diverging correlation lengths are observed on each side of the transition, and are related to the average size of force chain networks for $\phi<\phi_\mathrm{c}$~\cite{gajkinetic, gajthesis, gajfcorr1} and the average size of isostatic clusters for $\phi>\phi_\mathrm{c}$~\cite{floppy, silbertwyart}.  

In the jammed state many macroscopic observables exhibit power law scalings in $(\phi - \phi_\mathrm{c})$~\cite{ohern, zhangmakse} and the flowing state rheology changes dramatically as the packing fraction is increased above $\phi_\mathrm{c}$.  Although the jamming transition controls both dynamic and static properties near $\phi_\mathrm{c}$, theories tend to focus mainly on the latter~\cite{floppy, emt, criticalones}.  
In this Letter we theoretically explore the jamming transition by first studying flows with non-zero shear rate $\dot\gamma$ and then taking the limit of $\dot\gamma \rightarrow 0$ to access the static case.  This procedure leads to quantitative predictions for the flowing rheology that match observations in Ref.~\cite{campbellrigid}.  It also predicts the anomalous static scaling of the shear modulus measured in Ref.~\cite{ohern, zhangmakse} and a new scaling relation for the static yield strain, $\gamma^* \propto (\phi-\phi_\mathrm{c})^{1/2}$, which provides a testable prediction of the theory.

Granular materials behave as peculiar liquids.  This can be clearly demonstrated by measuring the stress tensor ${\bf \Sigma}$.  Depending on various parameters, the system will exhibit either Bagnold scaling ${\bf \Sigma} \propto \dot\gamma^2$, elastic-inertial scaling ${\bf \Sigma} \propto \dot \gamma^1$, or quasi-static scaling ${\bf \Sigma} \propto \dot\gamma^0$ (i.e. constant)~\cite{campbellrigid}.  The rheology of the flow is sensitively dependent on $\phi$, with Bagnold scaling for $\phi < \phi_\mathrm{c}$, quasi-static scaling for $\phi > \phi_c$, and elastic-inertial scaling in between.  This is in marked contrast to Newtonian fluids where ${\bf \Sigma} \propto \dot\gamma$ and only the proportionality constant depends on $\phi$.  Although the phase diagram of granular shear flow has been extensively studied in simulations~\cite{campbellrigid, othercampbellrigid, shearflowsim} and experiments~\cite{shearflowexp}, the origins of the rheological crossovers remain to be explained.
Arriving at a solution to this problem requires a well developed theory for the stress tensor in dense granular flows.

\begin{figure}
\begin{center}
\psfrag{pc}{\Huge{$\phi_\mathrm{c}$}}
\psfrag{phs}{\Huge{$\phi_\mathrm{hs}$}}
\psfrag{pqs}{\Huge{$\phi_\mathrm{qs}$}}
\psfrag{xicurve}{\Huge{$\xi$}}
\psfrag{lcurve }{\Huge{$l^*$}}
\psfrag{ctcurve }{\Huge{$c \tau$}}
\psfrag{sbg     }{\Huge{${\bf \Sigma} \propto \dot\gamma^2$}}
\psfrag{sei     }{\Huge{${\bf \Sigma} \propto \dot\gamma^1$}}
\psfrag{sqs     }{\Huge{${\bf \Sigma} \propto \dot\gamma^0$}}
\psfrag{packing }{\Huge{$\,\,\,\,\,\,\,\,\,\,\,\,{\bf \phi}$}}
%\scalebox{0.37}{\includegraphics{fig1.eps}}
\scalebox{0.36}{\includegraphics{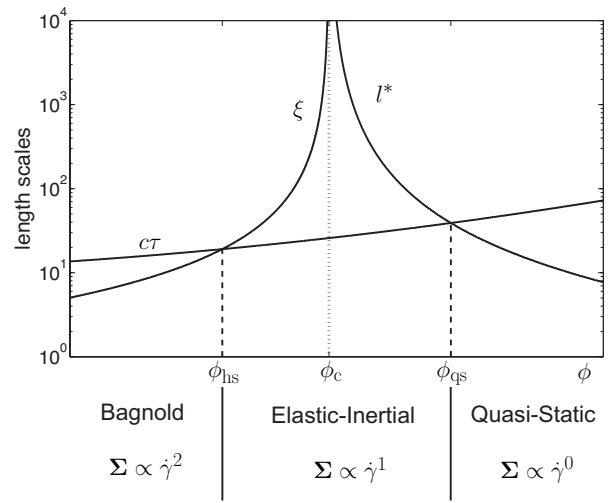}}
\vspace{-0.1in}
\caption{ \label{lengthfigure} A schematic illustration of three important length scales, plotted as a function of packing fraction $\phi$, and their relation to granular rheology near $\phi_\mathrm{c}$.  $\xi$ gives the size of force chain networks and diverges as $\phi \rightarrow \phi_\mathrm{c}^-$.  $l^*$ quantifies the size of marginally stable regions in the material and diverges as $\phi \rightarrow \phi_\mathrm{c}^+$.  Over length scales larger than $l^*$, the system responds as a conventional solid.  $c \tau$ represents the maximum correlation length through force networks and increases with packing fraction.  The macroscopic rheology (indicated beneath the graph) depends on the smallest length scale, and transitions occur at the intersections.  
%Our predictions for $\phi_\mathrm{hs}$ and $\phi_\mathrm{qs}$ are given in Eqns.~(\ref{criticalphiqs}, \ref{criticalphihs}).  
$\phi_\mathrm{c}$ is the packing fraction where the system jams in the limit of zero stress.  
Our predictions for $\phi_\mathrm{hs}$ and $\phi_\mathrm{qs}$ are given in Eqns.~(\ref{criticalphiqs}, \ref{criticalphihs}).  
The numerical values of the length scales depend on 
%the spatial dimension, friction coefficient, and other 
many parameters.  Here we have used $\xi = (\phi_c-\phi)^{-3/2}$, $\ell^* = (\phi-\phi_c)^{-2}$ and $c \tau = 10 + 30\phi^3$ for visualization, but the crucial features are the intersections rather than the specific numerical values.
}
\end{center}
\vspace{-0.3in}
\end{figure}

The stress tensor ultimately depends on the amorphous microscopic arrangement of grains.  Forces are transmitted via contacts between grains and a perturbation on one grain can have long range effects.  Indeed, both experiments and simulations indicate that contact forces are correlated~\cite{behringer, gajkinetic} and tend to form quasi one dimensional filaments, or force chains, that permeate the material~\cite{jgeng}.  These force chains assemble into network structures with a characteristic size $\xi$ that diverges as $\phi \rightarrow \phi_\mathrm{c}$ from below~\cite{gajkinetic, gajthesis,gajfcorr1}, as illustrated in Fig.~\ref{lengthfigure}.  Since contact between grains is the only form of force transfer, ${\bf \Sigma}$ is fully determined by properties of the networks~\cite{forcenet,gajthesis,gajfcorr2}.  Here we investigate the role of force chain networks in determining the value of the stress tensor by constructing a model of momentum transfer.

{\bf {\it The Force Network Model (FNM):}}
The central concept of the FNM is that the force $F^{ij}$ between a pair of contacting grains $\{i,j\}$ can be expressed as the sum of a collisional part $F^{ij}_\mathrm{bc}$ and a static part $F^{ij}_\mathrm{s}$.  The collisional part is proportional to the square of the relative velocity between grains $i$ and $j$ upon initial contact, is well described by kinetic theory~\cite{kineticreviews}, and scales with $\dot\gamma^2$.  It is the force that would be expected from collisions between pairs of grains.  The static force $F^{ij}_\mathrm{s}$ is a consequence of the pressure induced by the network surrounding a pair of grains $\{i,j\}$.  Collisional forces from other contacts in the network are transferred to $\{i,j\}$ from first nearest neighbors, second nearest neighbors, and so forth, all the way to the edge of the network.  The presence of multiple contacts in the cluster thereby increases the force at $\{i,j\}$ and this contribution can be expressed as a sum over all paths between the contact $\{i, j \}$ and every other contact $\{i',j'\}$ in the connected force network:  
\begin{equation}
F^{ij}_\mathrm{s} =  \sum_{\ell=1}^{\xi-1} N_\ell \mathcal{G}_\ell F^{i'j'}_\mathrm{bc}. 
\label{forcexfer}
\end{equation}
In this equation, $\ell$ is the path-length (in grain diameters) between the two contacts $\{i,j\}$ and $\{i',j'\}$, $N_\ell$ gives the number of pairs of contacts connected over a path of length $\ell$, and $F^{i'j'}_\mathrm{bc}$ is the collisional force on the contact at the end of the path.
The function $\mathcal{G}_\ell$ accounts for the fact that only a portion of the collisional force is transferred at each link in the path, and no more than the collisional force can be transferred from each contact in the network.  In constructing this equation we have assumed that grains are perfectly rigid, which ensures that the propagation of collisional forces is instantaneous.    

The FNM extends previous theories of granular flow that separate the stress tensor into static and collisional components~\cite{johnsonjackson} by introducing a natural physical basis for calculating the static stress.  Static stresses arise from clusters (not necessarily percolating) of simultaneously contacting grains, which exist even in the limit of infinitely rigid grains~\cite{gajkinetic,gajthesis, gajfcorr1}.  These force networks are created by dissipative contacts between grains and destroyed by the shear flow.  Since each of these rates is proportional to $\dot\gamma$ in the rigid grain limit, the properties of the steady state force networks are independent of $\dot\gamma$.  

{\bf {\it Bagnold Scaling:}}
Eqn.~(\ref{forcexfer}) yields a constitutive relation for each independent component of the stress tensor that does not contain any adjustable parameters and is expected to hold over scales much larger than the network size $\xi$.  This prediction has been shown to match data from simulations for all packing fractions $\phi < \phi_\mathrm{c}$~\cite{gajthesis,gajfcorr2}.  A major feature of the constitutive relations is that Bagnold scaling holds, or ${\bf \Sigma} \propto \dot\gamma^2$.  This is because all of the network parameters in Eqn.~(\ref{forcexfer}) are independent of $\dot\gamma$ and the rheology is thereby set by the average collisional force between grains, as predicted by kinetic theory~\cite{kineticreviews}.

The form of the constitutive relations is especially simple for rigid granular materials slightly below $\phi_\mathrm{c}$.
At large packing fractions, each grain has many contacts and the entire collisional force from every contact is transferred to its nearest neighbors.  Thus $\mathcal{G}_1$ equals its maximum value $1/N_1$, which ensures that $\mathcal{G}_\ell = 1/N_\ell$ for all values of $\ell$.  
In this limit, it follows from Eqn.~(\ref{forcexfer}) that 
%\begin{equation}
${\bf \Sigma} \sim \xi \langle F_{bc} \rangle \propto \xi \dot\gamma^2$.
%\label{stressscaling}
%\end{equation}
The stress tensor is predicted to depend linearly on the size of the force networks, independent of spatial dimension, and is insensitive to other network parameters.
      
{\bf {\it The Elastic-Inertial Regime:}}
The FNM prediction for the stress tensor has been constructed for perfectly rigid grains, but must be altered for realistic grains with a finite stiffness.  In particular, excitations do not propagate instantaneously through force networks, but occur at a finite speed $c$, which is an increasing function of the stiffness of the grains.  Combined with the average lifetime of the force chain networks $\tau$, the speed of force propagation defines a length scale $c \tau$ that gives the maximum extent forces can propagate through networks.    

If $\xi < c \tau$ then forces will propagate through the entire force network before it is destroyed.  Therefore the assumptions of the rigid grain FNM hold and ${\bf \Sigma} \propto \xi \dot\gamma^2$, averaged over time scales larger than $\xi/c$.
However, if $\xi > c \tau$, the collisional force from a single contact will not be transferred to the entire network, but only to contacts a distance $c \tau$ away.  Thus the network size $\xi$ is no longer relevant and the size $c \tau$ of the {\it correlated} networks should be used instead.  Since $\xi$ diverges at $\phi_\mathrm{c}$, there will always be a critical packing fraction $\phi_\mathrm{hs}$ above which $c \tau < \xi$, as illustrated in Fig.~\ref{lengthfigure}.  For $\phi > \phi_\mathrm{hs}$ the shear flow will be in a regime where $\xi$ must be replaced by $c \tau$.

The value of $c$ is a monotonically increasing function of the stiffness of grains and is also weakly dependent on the geometry of contact networks.  It is a property of the granular material and does not depend on the shear rate.  The average network lifetime $\tau$ is proportional to $\dot\gamma^{-1}$ since the shear flow pulls apart the networks~\cite{campbellrigid,othercampbellrigid}.  
% $\tau$ is not depend on the grain stiffness since it is a property of the {\it dynamic} contact networks, which are insensitive to the material properties of grains.  
Writing $\tau = \eta /\dot\gamma$ and substituting $\xi \rightarrow c \tau$ yields ${\bf \Sigma} \propto c \eta \dot\gamma$.  This predicts a linear scaling with $\dot\gamma$, as was observed in Ref.~\cite{campbellrigid} and named the elastic-inertial regime.  

The introduction of finite grain stiffness to the FNM produces a crossover from Bagnold's scaling to elastic-inertial scaling as the packing fraction or grain stiffness is increased.  The elastic-inertial scaling reveals that a second time scale, shown here to be proportional to the speed of force propagation through force networks, becomes important at high packing fraction.  

{\bf {\it The onset of quasi-static flow:}}            
%The value of $c \tau$ constrains the effective size of correlated force networks and controls the transition from the Bagnold regime to the elastic-inertial regime.  For granular shear flows at non-zero $\dot\gamma$, the stress tensor scales with $\dot\gamma$ as the system passes through the jamming transition.
%In dense regimes with $\phi > \phi_\mathrm{c}$, $\xi$ is no longer relevant since clusters regularly form that span the system.  Instead, $c \tau$ limits the extent of force propagation.
For granular materials with $\phi > \phi_\mathrm{c}$, another length scale $l^*$ becomes relevant~\cite{floppy} that is related to 
%dynamic modes and 
the departure of the material from the isostatic limit.  Isostatically jammed materials are configured such that contact forces can be determined identically from the constraint that no particle moves~\cite{isostaticorigin}.  This occurs when the coordination number $z$, equal to the average number of contacts per particle, approaches a critical value $z_\mathrm{c}$ that depends on the spatial dimension.  Contact forces are highly correlated in the isostatic state since breaking any contact will produce a cascade that alters the value of every other contact force.    
%and relates to the macroscopic behavior of the material.

The length scale $l^*$ quantifies the extent of cascades that result from contacts being broken.  For a system with coordination $z = z_\mathrm{c}+ \delta z$, a disturbance that breaks a single contact will result in a cascade over a length scale of $l^* \propto 1/\delta z $~\cite{floppy}, a scaling that has been verified in simulations~\cite{silbertwyart}.  The material is therefore fragile over length scales smaller than $l^*$, but the excess contacts $\delta z$ stabilize disturbances over length scales greater than $l^*$.  The value of $l^*$ is also related to the packing fraction, since $\delta z \propto (\phi - \phi_\mathrm{c})^{1/2}$~\cite{ohern, zhangmakse}, and is plotted in Fig.~\ref{lengthfigure}.

The presence of the length scale $l^*$
%, which only depends on geometric properties of the grain packing, 
has important dynamic effects when combined with the FNM and the length scale $c \tau$.
When $c \tau < l^*$, forces propagate over regions that are effectively isostatic and only marginally connected.  In this limit, if a contact is broken then there are rearrangements of grains over the entire range of the force chain networks and inertial scalings are important.  However, if $c \tau > l^*$, forces propagate over distances that are large compared to the rearranging regions, redundant contacts stabilize the flow, and forces are no longer inertial.  Thus, when $c \tau =l^*$ the system crosses over to a quasi-static regime with ${\bf \Sigma} \propto c^2/l^*$ for $c \tau \geq l^*$.  

{\bf {\it Predictions at non-zero shear rate:}}
The above considerations characterize the dependence of the stress tensor on the length scales $\xi$, $c \tau$, and $l^*$.  
This prediction is obtained using the FNM for rigid grains and carried through to the elastic-inertial and quasi-static regimes by investigating crossovers in the characteristic length scales.  The stress tensor is given by
\begin{equation}
{\bf \Sigma} \propto \Bigg\{ \begin{array}{cc} 
\xi \dot\gamma^2,& \mathrm{for} \, \, \xi \leq c \tau; \\
c \eta  \dot\gamma,& \mathrm{for} \, \, \xi \geq c \tau \, \, \mathrm{and} \, \, l^* \geq c \tau; \\
c^2 \eta^2 /l^*,& \mathrm{for} \, \, \xi \geq c \tau \, \, \mathrm{and} \, \, l^* \leq c \tau;\\
\end{array}
\label{fullstresstensor}
\end{equation}         
and a schematic of the various length scales and flow regimes is shown in Fig.~\ref{lengthfigure}.

\begin{figure}
\begin{center}
\psfrag{pcl}{\Huge{$\phi_\mathrm{c}$}}
\psfrag{yl}{\Huge{$\phi$}}
\psfrag{xl}{\Huge{$\dot\gamma^{\, \prime}$}}
\psfrag{phiqs }{\Huge{$\phi_\mathrm{qs}(\dot\gamma^{\, \prime})$}}
\psfrag{phihs }{\Huge{$\phi_\mathrm{hs}(\dot\gamma^{\, \prime})$}}
\scalebox{0.37}{\includegraphics{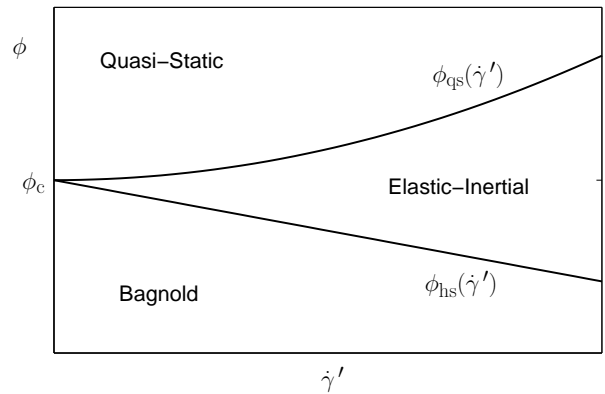}}
%\scalebox{0.30}{\includegraphics{fig2.eps}}
\vspace{-0.1in}
\caption{ \label{phasefigure} 
The predicted flow-map of granular materials, as a function of packing fraction $\phi$ and normalized shear rate $\dot\gamma^{\,\prime} = \dot\gamma/c\eta$.  The stress tensor scales with $\dot\gamma^2$ in the Bagnold regime, with $\dot\gamma^1$ in the elastic-inertial regime, and with $\dot\gamma^0$ in the quasi-static regime.  The scaling behavior of $\phi_\mathrm{qs}$ and $\phi_\mathrm{hs}$ are taken from Eqns.~(\ref{criticalphiqs},\ref{criticalphihs}), using $\psi=1$.   
}
\end{center}
\vspace{-0.3in}
\end{figure}

Instead of characterizing these regimes in terms of length scales, as is done in Eqn.~(\ref{fullstresstensor}), it is advantageous to recast the transition points in terms of a critical shear rate.
The transition from Bagnold to elastic-inertial scaling occurs when $\xi = c \tau$.  Given that $\tau =\eta/ \dot\gamma$, we can define a critical shear rate $\dot\gamma^*_\mathrm{hs}$.  Similarly, the transition from elastic-inertial scaling to quasi-static scaling occurs at a critical shear rate $\dot\gamma^*_\mathrm{qs}$.  
The critical values are
%\begin{eqnarray}
%\label{criticalgammas1}
%\dot\gamma^*_\mathrm{hs} &=& c \eta / \xi \\
%\dot\gamma^*_\mathrm{qs} &=& c \eta / l^*,
%\label{criticalgammas2}
%\end{eqnarray}  
\begin{equation}
\label{criticalgammas1}
\dot\gamma^*_\mathrm{hs} = c \eta / \xi, 
\vspace{-0.13 in}
\end{equation}
and
\vspace{-0.13 in}
\begin{equation}
\dot\gamma^*_\mathrm{qs} = c \eta / l^*.
\label{criticalgammas2}
\end{equation}  
Major features of the resulting granular rheology are illustrated in Fig.~\ref{phasefigure}.

For a granular shear flow with constant $\phi < \phi_\mathrm{c}$, corresponding to a horizontal slice through Fig.~\ref{phasefigure}, the system exhibits Bagnold scaling for $\dot \gamma < \dot \gamma^*_\mathrm{hs}$ and elastic-inertial scaling for $\dot \gamma > \dot \gamma^*_\mathrm{hs}$.  This unexpected behavior where Bagnold's scaling, normally associated with ``rapid'' flows, actually occurs for small $\dot\gamma$ in flows with constant packing has been observed previously~\cite{campbellrigid}.  For constant $\phi > \phi_\mathrm{c}$, the system exhibits quasi-static scaling for $\dot \gamma < \dot\gamma^*_\mathrm{qs}$ and elastic-inertial scaling for $\dot\gamma > \dot \gamma^*_\mathrm{qs}$.  The emergence of quasi-static flow as the shear rate is reduced in dense materials is a feature that has been observed in experiments~\cite{shearflowexp} and simulations~\cite{zhang1}.
%The stress tensor in the quasi-static regime is dependent on the wave speed of disturbances through force networks, and the isostatic length scale $l^*$.

The flow regimes can also be characterized by the packing fractions $\phi_\mathrm{hs}$ and $\phi_\mathrm{qs}$ at which the transitions occur.  For $c \tau \gg 1$, Fig.~\ref{lengthfigure} guarantees that the difference between the critical packing fractions and $\phi_\mathrm{c}$ will be small.  We can therefore use the scaling form $l^* \propto (\phi-\phi_\mathrm{c})^{-1/2}$ for the isostatic length scale in the dense regime~\cite{floppy, silbertwyart}.  
%The excess coordination scales as $\delta z \propto (\phi - \phi_\mathrm{c})^{1/2}$ for zero shear rate~\cite{ohern,zhangmakse}.  
%This form for the excess coordination should hold for $c \tau \gg 1$, which implies $\dot\gamma/c \eta \ll 1$.  
Combined with Eqn.~(\ref{criticalgammas2}), this predicts
\begin{equation}
\phi_\mathrm{qs} - \phi_\mathrm{c} \propto (\frac{\dot\gamma}{c\eta})^2.
\label{criticalphiqs}
\end{equation}
Thus as $\dot \gamma / c \eta $ approaches zero, the crossover to quasi-static flow occurs closer to $\phi_\mathrm{c}$.  
The functional form of $\phi_\mathrm{qs}$ is plotted in the flow-map of Fig.~\ref{phasefigure}.
%Moreover, the crossover is strongly dependent on the value of $\dot\gamma$.

We can also predict the value of $\phi_\mathrm{hs}$ in a similar way.  Since $\xi$ diverges at $\phi_\mathrm{c}$, then it must scale as $\xi \propto (\phi_\mathrm{c} - \phi)^{-\psi}$ near the transition.  For $c \tau \gg 1$ Eqn.~(\ref{criticalgammas1}) predicts
\begin{equation}
\phi_\mathrm{c} -\phi_\mathrm{hs} \propto \Big(\frac{\dot\gamma}{c \eta}\Big)^{1/\psi}.
\label{criticalphihs}
\end{equation}
%Although the value of $\psi$ has not been previously measured, an investigation of the crossover from Bagnold to elastic-inertial scaling in stiff granular materials provides a means to determine its value.
Eqns.~(\ref{criticalphiqs},\ref{criticalphihs}) fully determine the crossovers in rheology and allow us to construct the flow-map in Fig.~\ref{phasefigure}.  This figure is consistent with the numerically obtained flow-maps of granular rheology in Fig.~10 of Ref.~\cite{campbellrigid}, using $c \propto \sqrt{k}$, where $k$ is the grain stiffness.  
%In this case, $c \eta \propto \sqrt{k}$, so that the values of the crossover packing fractions can be related to the parameter $k \dot\gamma^{-2}$. 

{\bf {\it Predictions at zero shear rate:}}
In addition to predicting the rheology of granular materials, Eqn.~(\ref{fullstresstensor}) also gives the scaling of the stress tensor at zero shear rate when the system is jammed.  In the limit of $\dot\gamma \rightarrow 0$ the yield pressure $p_y$ and shear stress $s_y$ are both proportional to $c^2/l^*$.  For stresses less than these values, the system behaves elastically with $s = G \gamma$, where $G$ is the shear modulus and $\gamma$ is the shear strain.  We can therefore set $s_y = G \gamma^*$, with $\gamma^*$ the critical shear strain at which the system yields.  
%Because the material responds elastically, 
The speed of force propagation $c$ is then given by the speed of elastic shear waves:  $c \propto \sqrt{G}$.
%  In the case of simple shear that we investigate here, we may substitute $c$ by the transverse wave speed $c \propto \sqrt{G}$.

The dependence of $G$ and $p_y$ on packing fraction (or excess coordination) in the jammed state has been the focus of much attention.  The behavior of $p_y$ can be understood by assuming affine grain motion, whereas the scaling of $G$ is anomalous and affected by particle rearrangements after a strain step~\cite{ohern}.  Using $l^* \propto \delta z^{-1}$, Eqn.~(\ref{fullstresstensor}) predicts that
%\begin{eqnarray}
%G &\propto& p_y/ \delta z \\
%\gamma^* &\propto& \delta z.
%\label{zeroshearscaling}
%\end{eqnarray}
\begin{equation}
G \propto p_y/ \delta z, 
\label{Gscaling}
\vspace{-0.1 in}
\end{equation}
and
\vspace{-0.1 in}
\begin{equation}
\gamma^* \propto \delta z.
\label{zeroshearscaling}
\end{equation}
The first scaling relates the shear modulus to the yield pressure and excess coordination.  It matches measurements from simulation for all dimensions and interaction potentials~\cite{ohern,zhangmakse}.  The second relation predicts the critical shear strain at which the system yields.  Although this has not yet been measured in simulations or experiments, the FNM predicts that yielding is a purely geometric phenomena, unaffected by interaction potential.    

{\bf {\it Conclusions:}}
We have taken a dynamic approach to the jamming transition by using the FNM to predict the stress tensor in the flowing regime and then interpolating to the jammed state.  While this procedure was aimed at understanding athermal granular materials, a similar technique should apply in amorphous systems that jam as the temperature is reduced.  Since the isostatic length scale $\ell^*$ exists for all jammed amorphous materials, the challenge is to identify relevant correlation lengths in the flowing regime and relate them to properties of interest.  Jamming may then be universally understood as the transference of correlation from flowing length scales to the isostatic length scale.

{\bf {\it Acknowledgements:}}
%This work was supported by the William M. Keck Foundation, the MRSEC program of NSF under Award No. DMR00-80034, the James S. McDonnell Foundation, NSF Grant No. DMR-9813752, the David and Lucile Packard Foundation, and the NSF under Grant No. PHY99-07949.
This work was supported by the William M. Keck Foundation, the MRSEC program of NSF under Award No. DMR00-80034, the James S. McDonnell Foundation, the David and Lucile Packard Foundation, and NSF Grant Nos. DMR-9813752, PHY99-07949 and DMR-0606092.

\end{document}